# Sunsets, tall buildings, and the Earth's radius

**P.K.Aravind**
**Physics Department**
**Worcester Polytechnic Institute**
**Worcester, MA 01609**
**(paravind@wpi.edu)**

ABSTRACT

It is shown how repeated observations of the sunset from various points up a tall building can be used to determine the Earth's radius. The same observations can also be used, at some latitudes, to deduce an approximate value for the amount of atmospheric refraction at the horizon.

Eratosthenes appears to have been the first to have determined the Earth's radius by measuring the altitude of the sun at noon in Alexandria on a day when he knew it to be directly overhead in Syene (Aswan) to the south [1]. Despite the vastly better methods of determining the Earth's size and shape today, Eratosthenes' achievement continues to inspire schoolteachers and their students around the world. I can still recall my high school principal, Mr. Jack Gibson, telling us the story of Eratosthenes on numerous occasions in his inimitable style, which required frequent audience participation from the schoolboys gathered around him. In 2005, as part of the celebrations accompanying the World Year of Physics, the American Physical Society organized a project [2] in which students from 700 high schools all over North America collaborated in a recreation of Eratosthenes' historic experiment. The schools were teamed up in pairs, and each pair carried out measurements similar to those of Eratosthenes and pooled their results to obtain a value for the Earth's radius. The average of all the results so obtained yielded a value for the Earth's radius that differed from the true value by just 6%.

Over the years, a variety of elementary methods have been proposed for determining the Earth's radius [3-6]. It is the purpose of this paper to show how the method in [3] can be suitably extended and used in conjunction with a tall building to address this task. Student groups might find this an attractive project because it does not call for elaborate equipment and also provides them with the opportunity to journey to tall buildings in distant parts of the world.

The idea of the experiment is quite simple. One needs a tall building from which one can get an unobstructed view of the sunset over the ocean, at least after one has ascended a certain height up the building. The experiment consists of going up the building and observing sunset repeatedly from a succession of floors, and then using the times of the sunsets and the heights of the floors above sea level to deduce the Earth's radius. The idea of this experiment was laid out in [3], but without any mention of tall buildings. However the use of a tall building to carry out the experiment offers a distinct advantage, as will be pointed out below.

To understand the experiment in detail, it is necessary to know the formula linking the Earth's radius, $R$, to the time, $t_d$, by which sunset gets delayed when one rises a height $h$ above sea level. The formula is

$$1-\cos\left(\frac{2\pi}{T}t_d\right)=\frac{1}{\cos^2\lambda-\sin^2\delta}\frac{h}{R} \quad , \qquad (1)$$

where $\lambda$ is the latitude at which the building is located, $\delta$ is the sun's declination on the day of the experiment and $T$ (= 24hr = 86400s) is the Earth's rotational period. In (1), as well as in (2) below, $\lambda$ and $\delta$ are to be taken as positive or negative according as they are in the northern or southern hemisphere. Rawlins [3] quotes a formula equivalent to (1), but without derivation. I have given a derivation of (1) in Appendix 1 because it serves as the springboard for the derivation of the more complicated result (2) below. One sees that plotting the quantity on the left side of (1) as a function of $h$ yields a straight line whose slope can be used to determine the Earth's radius.

To get a feeling for the time delays one might encounter in practice, I took $\lambda = 40^\circ$, $\delta = 23.5^\circ$ and $R = 6378\text{km}$ in (1) and calculated $t_d$ for several different choices of $h$. The results are shown in the table below.

| Height $h$ (m) | 200 | 400 | 600 | 800 | 1000 |
|---|---|---|---|---|---|
| Time delay $t_d$ (s) | 166 | 235 | 288 | 332 | 372 |

If the declination is changed to $\delta = 15^\circ$ but all the other parameters are kept the same, the following table results:

| Height $h$ (m) | 200 | 400 | 600 | 800 | 1000 |
|---|---|---|---|---|---|
| Time delay $t_d$ (s) | 151 | 213 | 262 | 302 | 338 |

It is seen (at least for the above conditions) that there is generally less than a minute's difference between sunsets observed from points 200m apart in altitude, and that this interval shrinks as one goes upward. This might make it difficult for a single observer to carry out the experiment, but require a team of observers, with one observer stationed at each height. Of course, heights as large as those used in this example are not necessary. However, lower heights would lead to smaller differences in the sunset times at different heights and require more precise timing measurements to be made by the different observers.

It might be thought that determining the time delay, $t_d$, requires measuring the time of sunset at sea level, which may not be very practical. However this is unnecessary. The time of sunset at sea level can be looked up approximately in an almanac and fixed more precisely by the requirement that the plot of the left side of (1) against $h$ be a straight line passing through the origin. (To avoid any confusion, we should state that by "sunset" we mean the moment at which the sun's upper limb (and not its center) sets below the horizon. This definition is the most convenient one for the experiment considered here).

The determination of the Earth's radius using (1) requires only timing and altitude measurements, neither of which poses great challenges. The timing measurements can be made with the aid of accurate watches, with all the observers using synchronized watches. The altitudes of the floors above sea level can be determined with the aid of GPS or by other means. The latitude of the place can also be determined using GPS. While the determination of the altitudes should be straightforward, there are a variety of phenomena that could affect the timing measurements: the presence of ocean waves, unusual or variable atmospheric conditions [7], the absence of a sharp horizon, and poor visibility, to mention a few. Care should be taken to eliminate or minimize these effects, which might require repeating the experiment several times until reliable data are obtained.

The principal advantage of the present method over that in [3] is that multiple observations can be used to improve the accuracy of the result. The need to observe sunset at sea level, which is a considerable limitation, is also eliminated.

The main challenge in carrying out this experiment is finding a suitable building from which observations of the sunset can be made over the ocean (or over a large lake, such as Lake Michigan). An exotic alternative would be to observe the sunset from a balloon rising above the ocean. Alternatively, instead of sunset, one could observe the setting of a bright star such as Sirius.

There is one potentially serious effect that has been ignored completely in the above analysis – the effect of atmospheric refraction. Refraction actually causes the sun to be visible even after it has set entirely below the horizon. This is due to the fact that the "horizontal refraction" (i.e. the amount of refraction for an object at the horizon) is about $39'$ for Earth, whereas the diameter of the solar disk is about $30'$ [8]. An analysis of the effect of refraction on sunset times is carried out in Appendix 2 under the assumptions that (a) the amount of horizontal refraction, $\alpha_0$, is the same at all altitudes $h$, and (b) $\alpha_0$ and $\sqrt{h/R}$ are both small compared to unity and comparable in magnitude. Under these conditions, the formula that replaces (1) is

$$1-\cos\left(\frac{2\pi}{T}t_d\right)=\frac{1}{\cos^2\lambda-\sin^2\delta}\left(1+\frac{2\sin\lambda\sin\delta}{\cos^2\lambda-\sin^2\delta}\alpha_0\right)\frac{h}{R}+\frac{\sqrt{2}\sin\lambda\sin\delta}{\left(\cos^2\lambda-\sin^2\delta\right)^2}\left(\frac{h}{R}\right)^{3/2} , \quad (2)$$

where the second and third terms on the right (which are both of order $\alpha_0^3$) are the lowest order corrections to (1) arising from refraction and the finiteness of $h/R$. The terms that have been dropped in arriving at (2) are all of order $\alpha_0^4$ or higher. Introducing the notation $\Psi \equiv 2\pi t_d/T$, one sees from (2) that a plot of $\left(1-\cos\Psi\right)/h$ versus $\sqrt{h}$ yields a straight line whose slope can be used to determine $R$ and whose intercept with the vertical axis then determines $\alpha_0$. Again, a measurement of the time of sunset at sea level is unnecessary but can be fixed from the requirement that a straight line plot be obtained.

How serious are the effects of the two correction terms in (2)? The answer turns out to depend strongly on the conditions of the experiment. If the experiment is carried out at $\lambda = 40^\circ$, $\delta = 10^\circ$ and $h = 800\text{m}$ (and $\alpha_0 = 39' = .113$ rad), the last two terms in (2) add up to less than 1% of the main term and the effects of refraction are almost negligible. However if the experiment is carried out at $\lambda = 60^\circ$ and $\delta = 23^\circ$ (with the same values of $\alpha_0$ and $h$ as before), the correction terms add up to about 13% of the main term and the neglect of refraction would lead to a serious error; in this case $R$ and $\alpha_0$ should be determined in the manner explained in the previous paragraph.

An experiment that confirms the straight line plot predicted by Eq.(2) and uses it to infer the values of $R$ and $\alpha_0$ would be more challenging than a determination of $R$ based on Eq.(1). It would also require going to a location further north or south of the equator.

I hope to be able to carry out the above experiments with the aid of a team of students, but am not sure when that will happen. I am therefore publicizing the idea in the hope that another group might carry out the experiments and share the results with a wider audience. It would be interesting to see how good a value of the Earth's radius can be obtained by this variation of Eratosthenes' classic experiment.

**Acknowledgement**

I would like to thank Professor Adriaan Walther for a helpful discussion about the effects of atmospheric refraction.

**Endnote**

This paper is dedicated to the memory of John Travers Mends Gibson (1908-1994), principal of Mayo College, Ajmer, India, from 1954 to 1969, on the occasion of his birth centennial. He lives on fondly in the memories of many of us.

(Appendixes begin next page)

## Appendix 1: Derivation of Eq.(1)

Figure 1 shows the celestial sphere centered on the observer at O. The zenith, Z, is the point directly above the observer's head, and the observer's horizon is the plane through O perpendicular to the line OZ. The four cardinal points N,S,E and W are indicated on the observer's horizon. The observer is taken to be at latitude $\lambda$ in the northern hemisphere, and the polestar P is therefore at an angle $\lambda$ above his northern horizon. The sun, S', rises in the east and traces out a small circle on the celestial sphere as a result of the Earth's rotation. The sun's declination is taken to be $\delta$ north, and so the line from O to it sweeps out a cone with vertex O, axis OP and semi-vertex angle $(\pi/2-\delta)$ whose intersection with the celestial sphere gives the sun's path. The sun's path cuts the horizon at the sunrise point X and the sunset point Y.

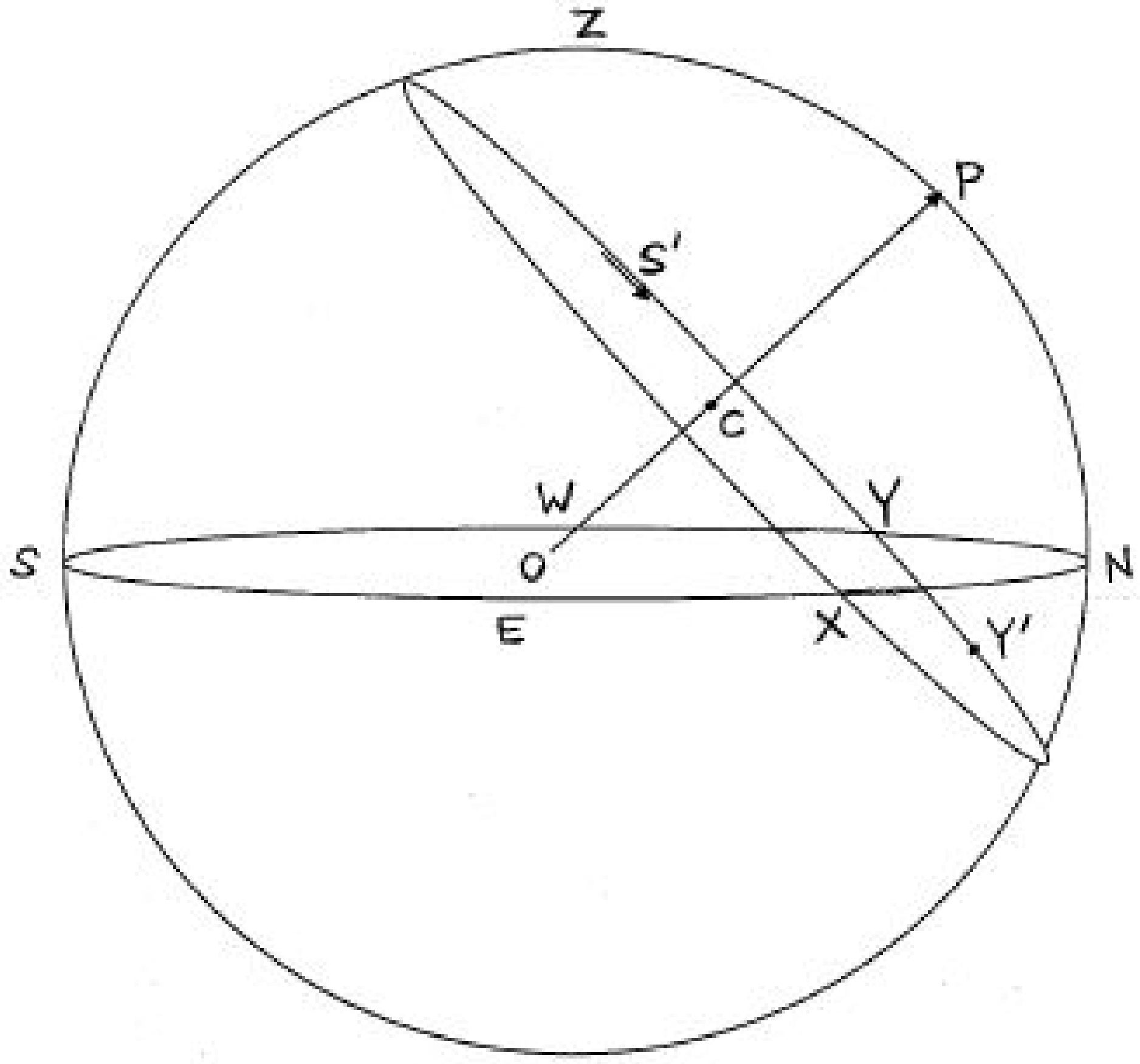

Fig.1. The celestial sphere, centered on the observer at O, with the sun's path shown on it as the slanted small circle. Some important angles in the figure are $\angle NOP = \lambda$, $\angle S'OP = \pi/2-\delta$, $\angle YOY' = \Phi$ and $\angle YCY' = \Psi$. See text for further explanation.

If the observer rises a height $h$ above sea level, s/he sits at the vertex of a cone that envelopes the Earth, as shown in Fig.2. The semi-vertex angle of this cone is $\alpha = \arcsin\left(R/\left(R+h\right)\right)$, where $R$ is the Earth's radius. Since $h$ and $R$ are both miniscule compared to the radius of the celestial sphere (which we take to be unity), we can take the vertex of this cone to be at O with negligible error. The surface of this cone intersects the celestial sphere in a small circle, the observer's new horizon, that is parallel to the earlier horizon and slightly below it. The sun's descent down the celestial sphere takes it from its initial setting point Y (when the observer is at sea level) to its later setting point Y' (when s/he is at height $h$ above sea level). Let C be the center of the small circle traced out by the sun on the celestial sphere and denote the angle YCY' by $\Psi$ (which we take to be in radians). The time between the sunsets at Y and Y' is then $t_d = \frac{\Psi}{2\pi} T$, where $T$ (= 24 hours) is the period of the sun's daily circuit on the celestial sphere. The task now facing us is to calculate $\Psi$ in terms of the quantities $\lambda, \delta, h$ and $R$, and thence to obtain an expression for $R$ in terms of the measured quantities $h$ and $t_d$ and the fixed parameters $\lambda, \delta$ and $T$.

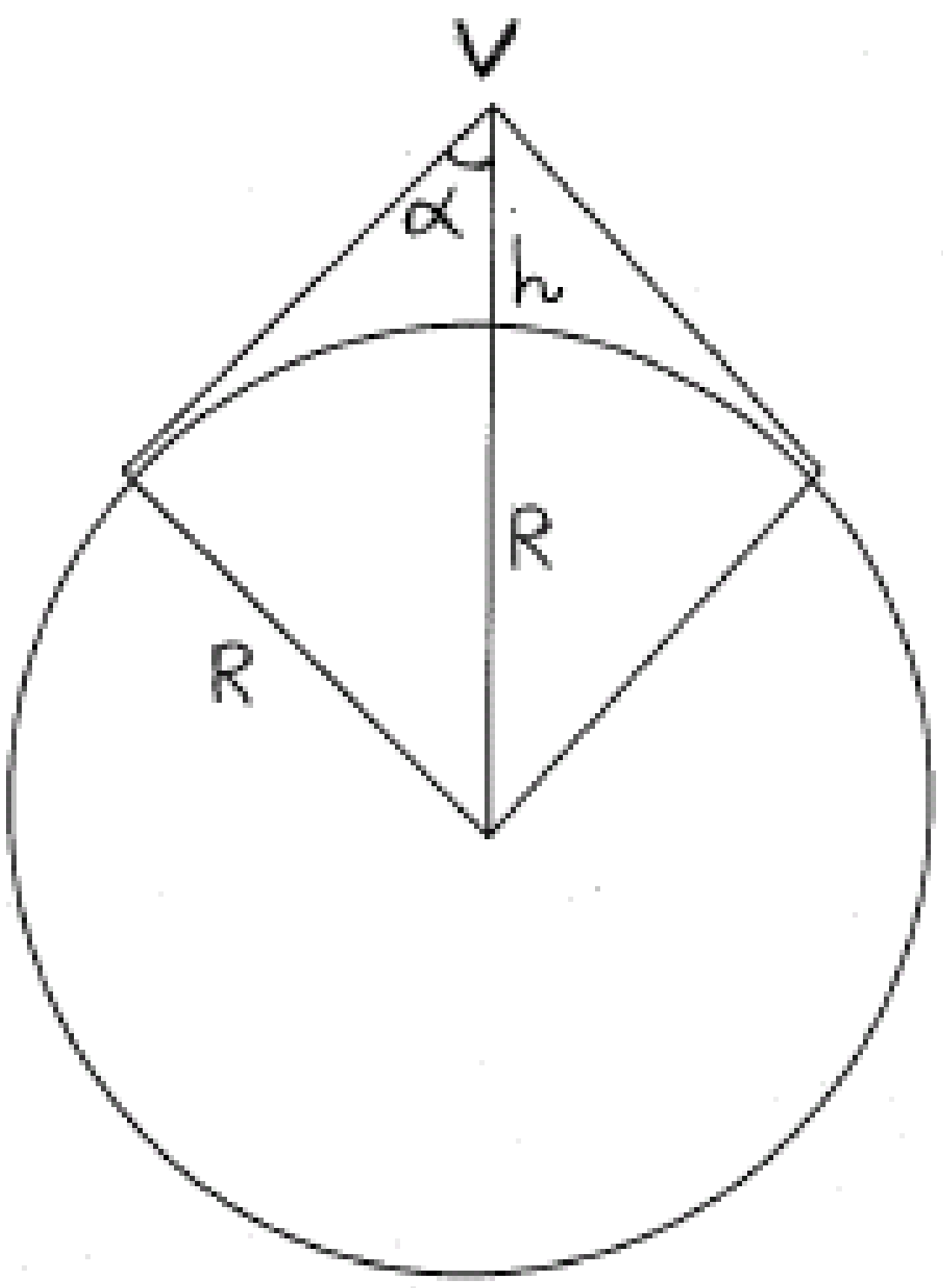

Fig.2. Showing the observer's horizon when s/he is at point V, a height $h$ above sea level. The horizon is the circle of tangency of the cone of vertex V and semi-vertex angle $\alpha = \arcsin\left(R / R+h\right)$ with the Earth (here $R$ is the Earth's radius). In most cases of practical interest, $h << R$ and $\alpha$ is only very slightly less than $90^\circ$.

Let us introduce a Cartesian coordinate system with its origin at O and its x-,y- and z-axes along OE, ON and OZ, respectively. Any vector from O to a point on the celestial sphere can be specified by its polar and azimuthal angles, the former being the angle between the vector and the positive z-axis and the latter the counterclockwise angle made by the projection of this vector on the x-y plane with the positive x-axis. Let $\vec{r}$ be the vector from O to the sun's position at any instant and let its polar and azimuthal angles be $\theta$ and $\phi$, respectively; then the Cartesian components of this vector are $\vec{r} = (\sin\theta\cos\phi, \sin\theta\sin\phi, \cos\theta)$. Let $\vec{n}$ be the vector from O to the polestar; it has polar angle $\pi/2-\lambda$ and azimuthal angle $\pi/2$ and hence the Cartesian components $\vec{n} = (0, \cos\lambda, \sin\lambda)$. The equation of the sun's path on the celestial sphere is $\vec{n}\cdot\vec{r} = \sin\delta$. The sunset point Y is one of the points of intersection of this path with the plane $z=0$ (the other being the sunrise point). Since the sunset point has polar angle $\pi/2$, its azimuthal angle can be found by putting $\theta = \pi/2$ in the equation of the sun's path and solving for $\phi$. This gives the azimuthal angle of Y as

$$\phi_1 = \arcsin\left(\frac{\sin\delta}{\cos\lambda}\right), \qquad (3)$$

where the obtuse angle solution is to be taken since the acute one corresponds to sunrise. The unit vector from O to the sunset point Y is then $\vec{v}_1 = (\cos\phi_1, \sin\phi_1, 0)$.

Next we determine the unit vector from O to the later sunset point Y'. Since the new horizon is a small circle on the celestial sphere with constant polar angle $\pi-\alpha$, one can determine the azimuthal angle of Y' by putting $\theta = \pi-\alpha$ in the equation of the sun's path and solving for $\phi$. This determines the azimuthal angle of Y' as

$$\phi_2 = \arcsin\left[\frac{\sin\delta + \sin\lambda\cos\alpha}{\cos\lambda\sin\alpha}\right], \qquad (4)$$

where again the obtuse angle solution is to be taken. The unit vector from O to the later sunset point Y' is then $\vec{v}_2 = (\sin\alpha\cos\phi_2, \sin\alpha\sin\phi_2, -\cos\alpha)$. The angle between the vectors $\vec{v}_1$ and $\vec{v}_2$, which we denote $\Phi$, is given by

$$\cos\Phi = \vec{v}_1\cdot\vec{v}_2 = \sin\alpha\cos(\phi_2-\phi_1). \qquad (5)$$

However the angle $\Psi$ that we want is the angle subtended by the points Y and Y' at C. Let $\vec{u}$ be the vector from O to C and $\vec{w}_1$ and $\vec{w}_2$ the vectors from C to Y and Y', respectively. Then $\vec{u} = (0, \sin\delta\cos\lambda, \sin\delta\sin\lambda)$, $\vec{w}_1 = \vec{v}_1 - \vec{u}$ and $\vec{w}_2 = \vec{v}_2 - \vec{u}$. The vector $\vec{u}$ has length $\sin\delta$, while $\vec{w}_1$ and $\vec{w}_2$ both have length $\cos\delta$. The relationship between $\Psi$ and $\Phi$ can be worked out by noting that

$$\vec{w}_1\cdot\vec{w}_2 = \cos^2\delta\cos\Psi = (\vec{v}_1-\vec{u})\cdot(\vec{v}_2-\vec{u}) = \vec{v}_1\cdot\vec{v}_2 - \vec{u}\cdot(\vec{v}_1+\vec{v}_2) + \vec{u}\cdot\vec{u} \qquad (6)$$

and using the relations $\vec{v}_1 \cdot \vec{v}_2 = \cos\Phi$ , $\vec{u} \cdot \vec{u} = \sin^2\delta$ and $\vec{u} \cdot \left(\vec{v}_1 + \vec{v}_2\right) = 2\sin^2\delta$ to find that

$$\cos\Psi = \left(\cos\Phi - \sin^2\delta\right) / \cos^2\delta \ . \tag{7}$$

Use of (3)-(5) then allows (7) to be recast as

$$\cos\Psi = \frac{c_1^{\ 2} + c_2\sqrt{1 - \left(2c_1 / c_2\right)\cos\alpha - \left(1/c_2\right)\cos^2\alpha} + c_1\cos\alpha}{c_3} \ , \tag{8}$$

where $c_1 = \sin\lambda\sin\delta, c_2 = \cos^2\lambda - \sin^2\delta$ and $c_3 = \cos^2\lambda\cos^2\delta$ . Recalling that $\Psi = 2\pi t_d / T$ and $\alpha = \arcsin\left(R/\left(R+h\right)\right)$, (8) is an implicit equation for $R$ in terms of $t_d, h, \lambda, \delta$ and $T$ . It cannot be used to obtain a closed form expression for $R$ , and one has to be content with a numerical solution instead.

The above solution is valid when $h$ has any magnitude relative to $R$ . The only approximation that has been made in arriving at this solution is that both $h$ and $R$ are vanishingly small compared to the radius of the celestial sphere (an excellent approximation indeed!). However, if one makes the further assumption that the ratio $x = h / R$ is small compared to unity, (8) can be simplified to give back (1). To see how this comes about, note that, for $x \ll 1$,

$$\cos\alpha = \frac{\sqrt{2x + x^2}}{1 + x} = \sqrt{2x} + \mathrm{O}\left(x^{3/2}\right) \tag{9a}$$

and $$\sqrt{1 - k_1\cos\alpha - k_2\cos^2\alpha} = 1 - \frac{1}{2}k_1\sqrt{2x} - \left(\frac{1}{4}k_1^{\ 2} + k_2\right)x + \mathrm{O}\left(x^{3/2}\right) \ , \tag{9b}$$

where $\mathrm{O}\left(x^n\right)$ denotes that the omitted terms are of order $x^n$ or higher. Using (9a) and (9b) in (8) leads to the series expansion

$$\cos\Psi = 1 - \frac{x}{\cos^2\lambda - \sin^2\delta} + \mathrm{O}\left(x^{3/2}\right) \ , \tag{10}$$

which, with the neglect of the $\mathrm{O}\left(x^n\right)$ terms, is (1).

In conclusion, it should be pointed out that (1) holds whether the sun's declination and/or the observer's latitude are in the northern or southern hemisphere. This can be seen by noting that $\lambda$ and $\delta$ get replaced by their negatives if they are in the southern hemisphere, and that this causes no change in (1). Finally we remark that the proposed experiment fails if one ventures so far north or south of the equator that the sun either never rises or sets in the course of a day, as happens when the denominator in (1) becomes negative.

## Appendix 2: Correction for refraction

Because of refraction, the sun, or any star, is visible even after it has set a certain angle $\alpha_0$ (the "horizontal refraction") below the horizon. This angle is on the order of $39'$ for the Earth. In our treatment of refraction we will assume that: (1) the horizontal refraction $\alpha_0$ is the same at all altitudes $h$, and (2) the dimensionless quantities $\alpha_0$ and $\sqrt{x}$ (where $x \equiv h/R$) are both much smaller than unity and of the same order of magnitude. Assumption (1) is not unreasonable and is made largely to simplify the analysis. Assumption (2) is well satisfied for a wide range of altitudes; as $h$ varies from 400m to 1km, $\sqrt{x}$ varies from .0080 to .0125, which is of the same order of magnitude as $\alpha_0 = .0113$ (the radian equivalent of $39'$).

We first work out the new sunset points at sea level and height $h$ in the presence of refraction. At sea level sunset occurs when the sun has set an angle $\alpha_0$ below the horizon, or when its polar angle is $\pi/2+\alpha_0$. Using this in the equation of the sun's path, $\vec{n}\cdot\vec{r} = \sin\delta$, allows one to solve for the azimuthal angle of sunset as

$$\phi_1^* = \arcsin\left(\frac{\sin\delta + \sin\lambda\sin\alpha_0}{\cos\lambda\cos\alpha_0}\right) \quad . \tag{11}$$

The first sunset point is then $\vec{v}_1^* = \left(\cos\alpha_0\cos\phi_1^*, \cos\alpha_0\sin\phi_1^*, -\sin\alpha_0\right)$, where $\phi_1^*$ is to be chosen in the second or third quadrant according as $\delta$ is in the northern or southern hemisphere. The sunset point at height $h$ above sea level can be found by setting the sun's polar angle equal to $\pi - \alpha + \alpha_0$ in the equation of its path and solving for the azimuthal angle as

$$\phi_2^* = \arcsin\left(\frac{\sin\delta + \sin\lambda\cos(\alpha-\alpha_0)}{\cos\lambda\sin(\alpha-\alpha_0)}\right) \quad . \tag{12}$$

The later sunset point is then at $\vec{v}_2^* = \left(\sin(\alpha-\alpha_0)\cos\phi_2^*, \sin(\alpha-\alpha_0)\sin\phi_2^*, -\cos(\alpha-\alpha_0)\right)$, with $\phi_2^*$ chosen in the same quadrant as $\phi_1^*$.

The angle $\Phi$ subtended by these new sunset points at the center of the celestial sphere is given by

$$\cos\Phi = \vec{v}_1^*\cdot\vec{v}_2^* = \cos\alpha_0\sin(\alpha-\alpha_0)\cos(\phi_2-\phi_1) + \sin\alpha_0\cos(\alpha-\alpha_0) \quad . \tag{13}$$

The angle $\Psi$ subtended by the sunset points at the center of the small circle in which the sun moves continues to be given by (6) and (7), with $\vec{u}$ as before but with $\vec{v}_1$ and $\vec{v}_2$ replaced by $\vec{v}_1^*$ and $\vec{v}_2^*$ and $\cos\Phi$ now given by (14). Using (11) and (12) in (13) to work out $\cos\Phi$ shows that $\cos\Psi$ is given in the present case by

$$\cos\Psi = \frac{1}{c_3}\left[c_2F_1F_2 + c_1^{\,2} + \sin\alpha_0\cos\left(\alpha-\alpha_0\right) + c_1\left\{\sin\alpha_0 + \cos\left(\alpha-\alpha_0\right)\right\}\right] \quad , \tag{14}$$

where $F_1 = \sqrt{1-\left(2c_1/c_2\right)\sin\alpha_0 - \left(1/c_2\right)\sin^2\alpha_0}$

and $F_2 = \sqrt{1-\left(2c_1/c_2\right)\cos\left(\alpha-\alpha_0\right) - \left(1/c_2\right)\cos^2\left(\alpha-\alpha_0\right)}$

and $c_1, c_2$ and $c_3$ were defined below Eq.(8) of Appendix 1.

We now expand $\cos\Psi$ as a power series in the small quantity $\sqrt{x}$ to find that

$$\cos\Psi = 1 + A_2x + A_3x^{3/2} + \mathrm{O}\left(x^2\right) \quad , \tag{15}$$

where $$A_2 = -\frac{\cos^2\alpha_0}{c_2 - 2c_1\sin\alpha_0 - \sin^2\alpha_0} \tag{16}$$

and $$A_3 = -\frac{\sqrt{2}\cos\alpha_0\left[c_1\left(1+\sin^2\alpha_0\right)+\left(1-c_2\right)\sin\alpha_0\right]}{c_2^{\,2} - 4c_1c_2\sin\alpha_0 + \left(4c_3 - 6c_2\right)\sin^2\alpha_0 + 4c_1\sin^3\alpha_0 + \sin^4\alpha_0} \quad . \tag{17}$$

Next we expand $A_2$ and $A_3$ as power series in the small quantity $\alpha_0$ to find that

$$A_2 = -\frac{1}{c_2} - \frac{2c_1}{c_2^{\,2}}\alpha_0 + \mathrm{O}\left(\alpha_0^{\,2}\right) \quad \text{and} \quad A_3 = -\frac{\sqrt{2}c_1}{c_2^{\,2}} + \mathrm{O}\left(\alpha_0\right) \quad . \tag{18}$$

Using (18) in (15) and doing a little rearrangement gives

$$1-\cos\Psi = \frac{1}{c_2}\left(1+\frac{2c_1}{c_2}\alpha_0\right)x + \frac{\sqrt{2}c_1}{c_2^{\,2}}x^{3/2} + \mathrm{O}\left(x^2\right) \quad , \tag{19}$$

where the neglected terms are of order $x^2$ or higher (taking $\alpha_0$ to be of order $\sqrt{x}$ in this count). Equation (19), with the $\mathrm{O}\left(x^2\right)$ terms neglected, is identical to Eq.(2) of the text.

If either $\lambda$ or $\delta$ are south of the equator, they should be taken with a negative sign in (2). This causes the second and third terms on the right of (2) to have positive signs if $\lambda$ and $\delta$ are in the same hemisphere and negative signs if they are in opposite hemispheres.